\documentclass[amssymb, preprintnumbers, showpacs, showkeys, aps, pra, superscriptaddress, twocolumn, longbibliography,10pt]{revtex4-1}

\usepackage[hidelinks, hypertexnames=false]{hyperref}   

\usepackage[all]{hypcap}
\usepackage{xcolor}
\usepackage{slashed}
\usepackage{graphicx}
\usepackage{amsmath}
\usepackage{latexsym}
\usepackage{epstopdf}
\usepackage{enumitem}
\usepackage{amssymb}
\usepackage{multirow}
\usepackage{mathtools}
\usepackage{bbm}
\usepackage{bm}
\usepackage{amsfonts}
\usepackage{amssymb}
\usepackage{calligra}
\usepackage{calrsfs}
\usepackage{makecell}
\usepackage[normalem]{ulem}
\usepackage[USenglish,american]{babel}
\usepackage{physics}
\usepackage{braket}
\usepackage{cancel} 

\newcommand{\be}{\begin{equation}}
\newcommand{\ee}{\end{equation}}
\newcommand{\ba}{\begin{eqnarray}}
\newcommand{\ea}{\end{eqnarray}}
\newcommand{\bs}{\begin{split}}
\newcommand{\es}{\end{split}}

\expandafter\ifx\csname package@font\endcsname\relax\else
 \expandafter\expandafter
 \expandafter\usepackage
 \expandafter\expandafter
 \expandafter{\csname package@font\endcsname}%
\fi

\begin{document}

\title{Supersolid dipolar phases in planar geometry: effects of tilted polarization.  
}

\author{Daniel Lima}%
\email{daniel.s.lima@posgrad.ufsc.br}
\affiliation{Departamento de F\'\i sica, Universidade Federal de Santa Catarina, 88040-900 Florian\'opolis, Brazil}

\author{Matheus Grossklags}%
\email{matheus.grossklags@posgrad.ufsc.br}
\affiliation{Departamento de F\'\i sica, Universidade Federal de Santa Catarina, 88040-900 Florian\'opolis, Brazil}

\author{Vinicius Zampronio}%
\email{v.zamproniopedroso@unifi.it}
\affiliation{Dipartimento di Fisica e Astronomia, Universit\`a di Firenze, I-50019, Sesto Fiorentino (FI), Italy}

\author{Fabio Cinti}%
\email{fabio.cinti@unifi.it}
\affiliation{Dipartimento di Fisica e Astronomia, Universit\`a di Firenze, I-50019, Sesto Fiorentino (FI), Italy}
\affiliation{INFN, Sezione di Firenze, I-50019, Sesto Fiorentino (FI), Italy}
\affiliation{Department of Physics, University of Johannesburg, P.O. Box 524, Auckland Park 2006, South Africa}

\author{Alejandro Mendoza-Coto}%
\email{alejandro.mendoza@ufsc.br}
\affiliation{Departamento de F\'\i sica, Universidade Federal de Santa Catarina, 88040-900 Florian\'opolis, Brazil}%
\affiliation{Dipartimento di Fisica e Astronomia, Universit\`a di Firenze, I-50019, Sesto Fiorentino (FI), Italy}

\begin{abstract}
The behavior of dipolar Bose-Einstein condensates in planar geometries is investigated, focusing on the effects of the polarization orientation. While perpendicular polarization produces a phase diagram with hexagonal, stripes, and honeycomb phases ending at a single critical point, the presence of an in-plane polarization component transforms the critical point into three critical lines, separating two phases at a time and changing radically the appearance of the phase diagram. All transition lines contain first- and second-order regions, while the phase diagram itself shows a resemblance with those displayed by quasi-one-dimensional dipolar systems. Finally, we investigate the effect of introducing an in-plane polarization on the structural properties of the phases and determine the superfluid fraction. Our results show that this process induces an axial deformation on the hexagonal and honeycomb phases, resulting in an anisotropic behavior in the long distance properties of the system like superfluidity. We expect that the rich phenomenology observed provides motivation for new experiments and theoretical works.
\end{abstract}

% The behavior of dipolar Bose-Einstein condensates in planar geometries is investigated focusing on the effects of polarization orientation. While perpendicular polarization produces a phase diagram with hexagonal, stripe, and honeycomb phases ending at an isolated critical point, the presence of an in-plane polarization component turns this critical point into three critical lines, separating two phases at a time and radically changing the appearance of the phase diagram. All transition lines contain first and second-order sectors while the phase diagram itself shows a resemblance with the one obtained for quasi-one-dimensional systems. Finally, the impact of introducing an in-plane polarization on the structural properties of the phases and the superfluid fraction is considered. Our results show that such a process induces an axial deformation in the system producing an anisotropic response, particularly in the superfluid behavior. We expect that the rich phenomenology observed serve as motivation for new experiments and theoretical works.

\maketitle
\section{Introduction}
The study of patterns with unconventional symmetries such as stripes phases~\cite{Berg_2007,Barci_2013,Mendoza_2015,Mendoza_2020_2}, smectic liquid crystals~\cite{Mendoza_2017,Radzihovsky_2020,Lahiri_2022}, cluster crystals~\cite{Prestipino_2014,Prestipino_2018,Mendoza_2021,Mendoza_2021_2,Mello_2023,Ciardi_2024}, and quasicrystals~\cite{Dotera2014,Barkan2014,Mendoza_2022,Ciardi_2023,Gross_2024,Zampronio_2024}, has become a central topic of modern many-body physics. These systems reveal a variety of phenomena over different physics fields, including soft matter \cite{Podoliak2023, Xia2024}, superconductivity \cite{BIANCONI, Bianconi2001}, cavity QED systems \cite{Mivehvar2021}, and long-range interacting systems \cite{Chomaz_2023}. Ultracold atomic systems have emerged as an ideal platform \cite{Mathey2009} to investigate such exotic states of matter, thanks to their unprecedented controllability and tunability \cite{Schafer2020}. In particular, dipolar BECs, characterized by long-range dipole-dipole interactions, have proven to be a quite versatile system in this context \cite{Chomaz_2023, Tanzi2019, Lu2011, Aikawa2012, Chomaz_2023}. Recent experiments have demonstrated the formation of quantum droplets and supersolids \cite{Tanzi2019, Chomaz2019, Bottcher2019}. In many cases the existence of such states can be seen as a macroscopic manifestation of quantum fluctuations \cite{Lee1957} since these effects stabilize phases that would otherwise undergo collapse due to the presence of attractive interactions \cite{Baillie2016, Bisset2016, Bombin2017}.

In the case of supersolids, a phase that simultaneously displays discrete translational symmetry and superfluidity \cite{Gross1957,Boninsegni2012}, initial experimental studies focused on quasi-one-dimensional systems, where density modulations are induced along an elongated trap \cite{Blakie2020}. However, recent breakthroughs have enabled the realization of two-dimensional supersolids \cite{Norcia2021, Bland2022}, producing new structural transitions and a large variety of crystalline phases. Theoretical investigations~\cite{Lu2015, zhang2019, Zhang2023, Poli2021, Hertkorn2021, Schmidt2022} have predicted complex phase diagrams that include hexagonal, stripes and honeycomb configurations, with phase transitions governed by density and interaction parameters \cite{Zhang2021, Ripley2023}. Notably, the emergence of metastable states, such as ring-lattice patterns, highlights the intricate competition between different symmetries and interactions~\cite{Zhang_2024}.

Recently, a series of works \cite{zhang2019, Ripley2023,Zhang_2024, Kora2019,PhysRevA.96.013627} have addressed in detail the study of the ground-state phase diagram of a Bose dipolar gas in planar geometry when the system is polarized perpendicular to its plane. Those works have shown that, in such conditions, the system develops three modulated phases: a hexagonal solid at low densities; a stripes phase at intermediate densities; and a honeycomb phase at high densities. The transition between these phases is always first order, except at the critical point of the system in which all phase boundaries converge. The existence of this critical point in the thermodynamic limit is particularly interesting considering that soft-core bosonic supersolid phases usually display a first-order transition between the homogeneous and the modulated phases \cite{Pohl2013, Kunimi2012, Hsueh2012, Saccani2012}. Moreover, from the experimental point of view, the existence of a continuous supersolid-superfluid transition could facilitate the production of the 2D supersolid phase \cite{Bland2022}, since typically metastable states are produced when the system is driven from the superfluid to the supersolid phase.

\begin{figure}[!t]
    \centering
    \includegraphics[scale=0.9]{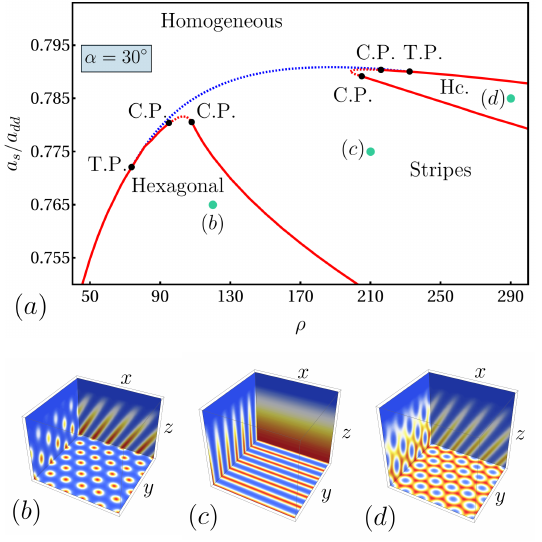}
    \caption{Phase diagram and modulated phases developed by a dipolar planar bosonic gas for a polarization orientation $\alpha=30^{\circ}$, with respect to the normal direction to the plane of the system. In (a), we present the phase diagram of the system at a fixed polarization orientation varying the average density of particles $\rho$ and the ratio $a_s/a_{dd}$. In (b-d) we show slice density plots of the 3D density patterns exhibited by the system in the modulated phases identified.  The areal average density of particles is measured along the plane perpendicular to the polarization vector.}
    \label{Fig1}
\end{figure}

In this work, we study the effects of arbitrary polarization orientation for a dipolar boson system in planar geometry. Our results show that when the polarization vector exhibits a component along the system's plane, the critical point predicted for the case of perpendicular polarization evolves into multiple critical lines. As an example, in Fig.~\ref{Fig1}(a), we show the phase diagram obtained for a tilting angle $\alpha=30^{\circ}$. We were able to identify three different modulated phases according to the density pattern in the $xy$-plane of the system: a compressed hexagonal lattice, a stripes phase, and a stretched honeycomb phase (see Fig.~\ref{Fig1}(b-d)). As can be observed, the topology of the phase diagram obtained differs significantly from the one known for perpendicular polarization, displaying some similarities with the phase diagram of quasi one-dimensional cigar-shaped dipolar systems. Indeed, as can be observed, the system presents a wide intermediate density region in which the transition from the one-dimensional modulated pattern (stripes) to the homogeneous superfluid phase is second order. Moreover, we investigate how the polarization orientation impacts the structural properties of the ground-state phases and the superfluid fraction.

% Our results show that when the polarization vector of the system presents a component along the system's plane, the critical point predicted for the case of perpendicular polarization evolves into various critical lines. As an example, in Fig.\ref{Fig1}(d) we show the phase diagram obtained for a tilting angle $\alpha=30^{\circ}$. We were able to identify three different modulated phases according to the density pattern in the $xy$-plane of the system, a compressed hexagonal lattice, a stripes phase, and a stretched honeycomb phase (see Fig.\ref{Fig1}(a-c)). As can be observed the topology of the phase diagram obtained differs significantly from the one known for perpendicular polarization, presenting some resemblance with the phase diagram obtained for one-dimensional cigar shape dipolar systems. Indeed, as in the quasi 1D scenario, the system presents a wide intermediate density region at which the transition from the one-dimensional modulated solution (stripes) to the homogeneous superfluid phase is second order. Moreover, we investigate how the polarization orientation impacts the structural properties of the ground state phases and the superfluid fraction

This paper is organized as follows: Sec. \ref{section:model_and_methods} introduces the microscopic model and outlines the theoretical framework used to describe the modulated phases of a BEC in planar geometry with tilted polarization. Sec. \ref{single_mode_results} is dedicated to presenting the evolution of the phase diagram of the single-mode system as the tilting polarization angle is varied. Furthermore, Sec. \ref{section:many_mode_results} focuses on the study of the properties of the system using fully converged spectral expansions. Here, we present a comparison with single-mode results and investigate in detail the structural properties of the system as well as the impact of the tilted polarization on the superfluid properties. Finally, Section V delivers an ending discussion and concluding remarks.

\section{Model and Methods}\label{section:model_and_methods}
%(Apresentar o sistema, inclusive matematicamente, possivelmente incluindo figuras que esclareçam. %Chegar no funcional energia por particula. Apresentar os anzats que serao considerados para cada tipo %de estado.)

We consider a three-dimensional system of $N$ identical dipolar bosons with mass $m$ at zero temperature, confined by a harmonic trap oriented along the direction $z$ and free in the $xy$-plane (see Fig.~\ref{Fig1}). The particles interact through collisions, modeled as a zero-range interaction with scattering length $a_s$ as well as through a non-local dipolar interaction, characterized by a scattering length $a_{dd}$. Moreover, we consider that the magnetic moment of all particles is polarized along the direction $(\sin(\alpha),0,\cos(\alpha))$, i.e., forming an angle $\alpha$ with the trapping direction. For simplicity in the mathematical description, we work in a rotated coordinate system of the form  $x'=x\cos(\alpha)-z\sin(\alpha)$, $y'=y$ and $z'=z\cos(\alpha)+x\sin(\alpha)$. In this new coordinate system, the polarization direction coincides with the $z'$ axis and the central plane of the system is given by $z'=\tan(\alpha)x'$. To avoid cumbersome notation, from now on we will omit the prime symbol in our calculations to refer to the new coordinate system. Taking as units of length $\delta l=12\pi a_{dd}$ and time $\delta t=m\delta l^2/\hbar$, respectively, the mean-field energy per particle functional can be expressed as  
\begin{align}\label{EN}
&\frac{E}{N}=\int d\textbf{x}\left\{\frac{1}{2}\vert\bm{\nabla}\psi\vert^2+U(x,z)\vert\psi\vert^2+\frac{2}{5}N^{3/2}\gamma\vert\psi\vert^{5/2}\right.\nonumber\\
+&\left.N\frac{a_s}{6a_{dd}}\vert\psi\vert^4+\frac{N}{2}\int d\textbf{x}'V(\textbf{x}-\textbf{x}')\vert\psi(\textbf{x})\vert^2\psi(\textbf{x}')\vert^2\right\}.
\end{align}
The trapping potential is modeled as $U(x,z)=\frac{1}{2}\omega^2(z\cos(\alpha)-x\sin(\alpha))^2$, where $\omega$ stands for the trap frequency, while the Lee-Huang-Yang (LHY) \cite{Lee1957} coefficient can be written as $\gamma=\frac{4}{3\pi^2}\left(a_s/(3a_{dd})\right)^{5/2}\left[1+\frac{3}{2}(a_{dd}/a_s)^2\right]$ \cite{Lima2011,Lima2013}. Finally, the dipole-dipole interaction potential $V(\textbf{r})$ for a system polarized along the $z$-direction is given by 
\begin{align}
V(\textbf{r})=\frac{1}{r^3}\left(1-\frac{3z^2}{r^2}\right).
\end{align}
The ground-state phase diagram of the model in the thermodynamic limit can be accessed by minimizing the energy per particle functional subject to the constraint $\int d\textbf{x}\vert\psi(\textbf{x})\vert^2=1$. Here, the integration domain extends over the region $[0,L_x]\times [0,L_y]$, such that $L_x\times L_y\rightarrow\infty$ while $z\in(-\infty,+\infty)$. Moreover, the number of particles in the system is given by $N=\rho L_xL_y$, where $\rho$ stands for the areal density of particles along the $xy$-plane perpendicular to polarization direction~\footnote{Here is important to notice that the areal density along the plane of the system ($\rho_\perp$) set by the harmonic trap can be written as $\rho_\perp=\rho \cos(\alpha)$}. 

% In agreement with the obtained results for the situation in which the magnetization is oriented perpendicular to the plane of the system ($\alpha=0$), we consider that in our case the ground state wave function of the system obeys the ansatz

In agreement with the scenario in which the magnetization is oriented perpendicular to the plane of the system ($\alpha=0$), we consider a ground state wave function of the form
\begin{equation}
\psi(\textbf{x})=\frac{1}{\sqrt{A}}\phi(x,y)\chi(z_\perp(x,z)),
\end{equation}
where $A=L_xL_y$ stands for the area of the system, $z_\perp=z\cos(\alpha)-x\sin(\alpha)$, and $\chi^2(z_\perp(x,z))$  describes the localized density profile produced by the confining potential $U(z_\perp(x,z))$ on the transversal direction (see Fig.~\ref{Fig1}). Without  generality loss, we replace the normalization condition for $\psi(\textbf{x})$ by the following two independent normalization conditions for $\chi(z_\perp)$ and $\phi(x,y)$, $\int dz\chi^2(z_\perp(x,z))=1$ and $\int_A dx dy \vert\phi\vert^2(x,y)=A$. In analogy with previous works, we consider that $\chi^2(z_\perp)$ follows a Thomas-Fermi profile of the form 
\begin{equation}
\chi^2(z_\perp)=\frac{3\cos(\alpha)}{4\sigma}\left(1-\frac{(z_\perp)^2}{\sigma^2}\right)\Theta\left(\sigma-\vert z_\perp\vert\right),
\end{equation}
where $\sigma$ represent the width of the system to be determined variationally and $\Theta(u)$ stand for the Heaviside step function. Replacing the proposed ansatz for $\psi(\textbf{x})$ in Eq.~\eqref{EN}, we obtain that the energy per particle of the system can be rewritten as  
\begin{align}
\nonumber
\frac{E}{N}=&\int \frac{d\textbf{r}}{A}\left\{\frac{1}{2}\vert\bm{\nabla}\phi\vert^2+\frac{\omega^2\sigma^2}{10}\right.\\ \nonumber
+&\left.\frac{9\sqrt{3}\pi}{256}\frac{a_s}{a_{dd}}\frac{\gamma(\rho\cos(\alpha))^{3/2}}{\sigma^{3/2}}\vert\phi(\textbf{r})\vert^{5}+\frac{\rho\cos(\alpha)}{10\sigma}\frac{a_s}{a_{dd}}\vert\phi(\textbf{r})\vert^4\right.\\ 
+&\left.\frac{\rho \cos(\alpha)}{2\sigma}\int d\textbf{r}'V_{\mathrm{eff}}(\textbf{r}-\textbf{r}')\vert\phi(\textbf{r})\vert^2\vert\phi(\textbf{r}')\vert^2\right\},
\label{EN2}
\end{align}
where $\textbf{r}$ stands for the vector position in the $xy$-plane and the effective potential $V_{\mathrm{eff}}(\textbf{r})$ is defined by its form in momentum space $\hat{V}_{\mathrm{eff}}(\textbf{k})=2/5-$$F\left(k_x\sigma/\cos(\alpha),k_y\sigma/\cos(\alpha),\alpha\right)$ where 
\begin{align}
\nonumber
F(q_x,q_y,\alpha)=&\int_{-\infty}^{\infty} \frac{dq}{2\pi}\left(\frac{3(-q\cos(q)+\sin(q))}{q^3}\right)^2\\
\times&\frac{q_y^2+(q_x-q\tan(\alpha))^2}{q_y^2+q^2+(q_x-q\tan(\alpha))^2}.
\end{align}
It is worth mentioning that in Eq.~\eqref{EN2} the kinetic energy contribution along the $z$-direction has been neglected as usual within the Thomas-Fermi approximation. This procedure has been validated by previous works reporting similar results using or not such an approximation \cite{zhang2019, Zhang2021}.

Now that the energy per particle functional for the 2D effective problem associated with the configurations of $\phi(\textbf{r})$ have been constructed, the ground-state properties of the system, including its phase diagram, can be determined from the direct minimization of such functional. With this goal, we consider three different kinds of possible solutions: a homogeneous state, a stripes configuration along the $x$-axis, and a 2D solution with hexagonal symmetry that can be stretched or compressed along the $x$-axis as well. The $x$-axis, in our case, coincides with the direction of the in-plane polarization of the system.

\subsection{Homogeneous state}
In the case of the homogeneous state $\phi_h(\textbf{r})=1$, implying an energy per particle of the form
\begin{align}
\nonumber
\frac{E_h}{N}=&\frac{\omega^2\sigma^2}{10}+\frac{9\sqrt{3}\pi}{256}\frac{\gamma(\rho\cos(\alpha))^{3/2}}{\sigma^{3/2}}+\frac{\rho\cos(\alpha)}{10\sigma}\frac{a_s}{a_{dd}}\\
+&\frac{\rho\cos(\alpha)}{5\sigma}\left(1-\frac{3}{2}\sin(\alpha)^2\right).
\end{align}
The properties of this phase are then obtained after the minimization of $E_h/N$ with respect to the single parameter $\sigma$. Moreover, we verified that the value of $\sigma$ for all the modulated solutions that will be presented shortly does not differ significantly from its corresponding value for the homogeneous solution. For this reason, to simplify the numerical evaluation, we will approximate the value of $\sigma$ in those cases by the corresponding value obtained for the homogeneous solution.

\subsection{Stripes state}
In this case, we consider that the ground state wave function takes the form 
\begin{align}
\phi_{\mathrm{st}}(\textbf{r})=\frac{1+\sum_{n\geq1}c_n\cos(n k_0y)}{1+\frac{1}{2}\sum_{n\geq1}c_n^2},
\label{stripes}
\end{align}
where the set of coefficients $c_n$'s are taken as variational parameters as well as the wave vector of the modulation $k_0$. The normalization factor introduced in the denominator of the right-hand side of Eq.~\eqref{stripes} is required to guarantee the normalization condition previously mentioned for $\phi(\textbf{r})$. An important aspect to be noticed is that we have explicitly considered that stripes are oriented along the in-plane component of the polarization vector, which in our case coincide with the $x$-axis. This alignment is produced by the attractive component of the dipolar interaction that creates an anisotropy minimizing the energy in such configuration.

% where the set of coefficients $c_n$'s are taken as variational parameters as well as the wave vector of the modulation $k_0$. We should notice that the normalization factor included in the denominator of the {\color{red}right-hand side} of Eq.~\eqref{stripes} is required to guarantee the normalization condition for $\phi(\textbf{r})$ previously mentioned. An important aspect to be noticed is that we have explicitly considered that stripes are oriented along the $x$-axis. This is justified since, in the case of a tilted polarization, the dipolar interaction produces an anisotropy in the system that favors the alignment of the stripes with the projection of the polarization over the central plane of the harmonic trap. 

\subsection{Compressed and Stretched hexagonal state}

In analogy with the results obtained for dipolar gases confined in a planar geometry, we also consider compressed and stretched solutions with hexagonal symmetry of the form
\begin{align}
\phi_{\mathrm{hx}}(\textbf{r})=\frac{1+\sum_{m,n}^{'}\frac{c_{m,n}}{2}\cos(k_0\textbf{e}_{nm}\cdot\textbf{r})}{\left(1+\sum_{m,n}^{'}\frac{c_{n,m}^2}{4}\right)},
\label{bubbles}
\end{align}
where $\textbf{e}_{m,n}=m\textbf{e}_1+n\textbf{e}_2$, with $\textbf{e}_1=(0,1)$ and $\textbf{e}_2=(1/2\cot(\theta),-1/2)$ and $m$ and $n$ integers. The prime on top of the sum symbol excludes the case $(m,n)=(0,0)$. The form of the selected basis $\{\textbf{e}_1,\textbf{e}_2\}$ is general enough to allow simultaneously homogeneous axial deformations of the hexagonal pattern along the $x$-axis and to recover continuously the energy functional corresponding to the fully symmetric hexagonal pattern, when we have $\alpha=0$ and $\theta=\pi/6$. It is worth noticing that a solution with $\theta<(>)\ \pi/6$ corresponds to a hexagonal pattern compressed (stretched) along the $x$-axis. Moreover, the variational coefficients $c_{n,m}$'s are set to be equal when they correspond to equivalent wave vectors $k_0\textbf{e}_{m,n}$ by inversion and reflection symmetries with respect to the $x$ and $y$ axes, since these symmetries will be preserved by a uniform axial deformation along the $x$-direction.
Finally, it is worth noticing that in this case, besides the variational parameters $c_{n,m}$'s, the characteristic wave vector $k_0$ and the parameter $\theta$ are determined from the minimization of the energy per particle once the anzats in Eq.~\eqref{bubbles} is inserted in the energy functional in Eq.~\eqref{EN2}.      
\begin{figure*}[!t]
    \centering
    \includegraphics[width=1.00\textwidth]{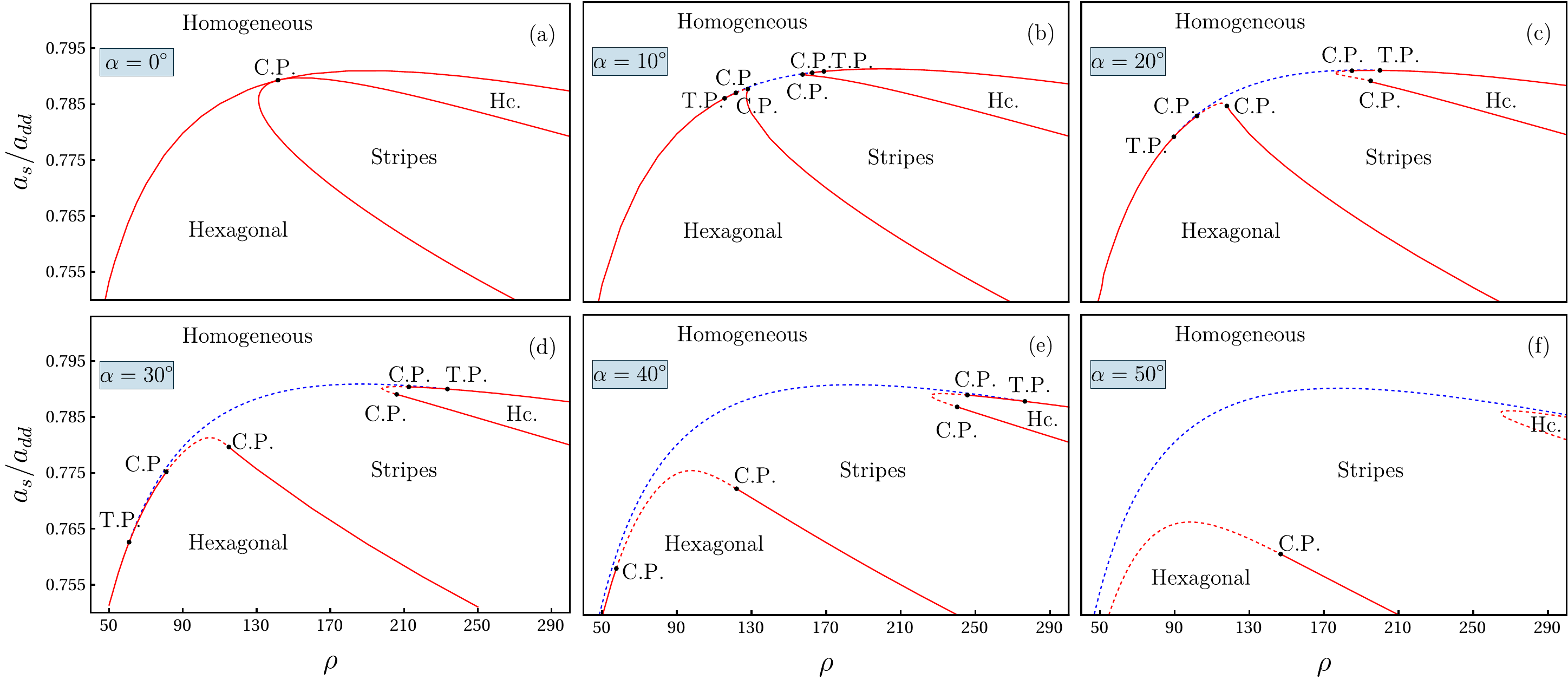}
        \caption{Ground state phase diagrams for a fixed trap frequency $\omega=0.08$ for different values of the tilting angle $\alpha$. Four distinct phases are observed: a homogeneous phase, a hexagonal compressed phase, a honeycomb stretched (Hc.) phase, and a stripes phase. Dashed lines correspond to continuous transition, while full lines correspond to first-order transitions between these phases. The acronyms C.P. and T.P. stand for critical point and triple point, respectively.}
    \label{PhaseDiags}
\end{figure*}

\subsection{Variational method}
Considering that the actual ground-state wave function $\phi(\textbf{r})$ has the symmetries of a homogeneous, stripes or compressed (stretched) hexagonal state, the projection of the original problem to the momentum (Fourier) space is a completely general transformation that turns the original variational problem into one of minimizing a many variable function. The main advantage of such a method over standard finite difference methods is related to the fact that, unless we go deep into the modulated regions of the phase diagram, the number of relevant Fourier amplitudes is usually quite limited, improving significantly the computational cost of the variational problem. Moreover, since the energy per particle functional is polynomial in $\phi(\textbf{r})$, an exact evaluation in terms of Fourier amplitudes is possible for all the modulated phases upon spatial integration.   

% Considering that the actual ground-state wave function $\phi(\textbf{r})$ has the symmetries of a homogeneous, stripes or {\color{red}compressed (stretched)} hexagonal state, the projection of the original problem to the momentum (Fourier) space is a completely general transformation that turns the original variational problem into one of minimizing a many variable function. The main advantage of such a method over standard finite difference methods is related to the fact that, unless we go very deep into the modulated regions of the phase diagram, the number of Fourier amplitudes taking significant values is usually quite limited, improving significantly the computational cost of the variational problem. Moreover, since the energy per particle functional is polynomial in $\phi(\textbf{r})$, an exact evaluation in terms of Fourier amplitudes is possible for all the modulated phases upon spatial integration.

\subsection{Superfluid tensor}
As discussed in the literature, the superfluid fraction tensor for a supersolid of mass $M$ can be defined from the reduction of its translational inertia when we consider that the system is confined by moving walls. Thus, the superfluid tensor is defined as 
\begin{equation}
f_{s}^{ij}=\delta_{ij}-\lim_{v\rightarrow0}\frac{1}{M}\frac{\partial P_i}{\partial v_j}, \ \ \ \ i,j=\{x,y\}
\end{equation}
where $v_j$ represent the $j$-component of the moving walls and $P_i$ represent the equilibrium momentum of the system. Considering the problem of the equilibrium state of the system with such boundary conditions, \textcolor{blue}{it} is possible to conclude that the superfluid fraction tensor can be operationally computed as \cite{Blakie2024, Josserand2007, Sepúlveda2010}\begin{equation}
f_{s}^{ij}=\delta_{ij}-\int \frac{d\textbf{r}}{A}\vert\phi(\textbf{r})\vert^2\frac{\partial K_j}{\partial x_i},
\end{equation}   
where the auxiliary functions $K_x(\textbf{r})$ and $K_y(\textbf{r})$ satisfies respectively the equations 
\begin{equation}
\bm{\nabla}\cdot\left(\rho(\textbf{r})\bm{\nabla}K_i\right)=\frac{\partial \rho(\textbf{r})}{\partial x_i}, 
\label{EqK}
\end{equation}
where $\rho(\textbf{r})=\vert\phi(x\cos(\alpha),y)\vert^2$. This expression for $\rho(\textbf{r})$ corresponds to the probability density along the central plane of our system, which is the one to be considered if we want to study its planar-superfluid properties.

To proceed, instead of directly solving the problem posed by Eq.~\eqref{EqK}, we took a different route to determine $K_{x,y}(\textbf{r})$. Our approach is based on recognizing that Eq.~\eqref{EqK} corresponds to the Euler-Lagrange equation of the action
\begin{equation}
S[K_i]=\int \frac{d\textbf{r}}{A}\left(K_i(\textbf{r})\frac{\partial \rho(\textbf{r})}{\partial x_i}+\frac{\rho(\textbf{r})}{2}(\bm{\nabla}K_i)^2\right).
\label{action}
\end{equation}
It is not difficult to notice that the above functional for $i={x,y}$ is convex, hence the solutions of Eq.~\eqref{EqK} correspond to their respective global minimum. We should notice that since the form of $\rho(\textbf{r})$ is known, the symmetry properties and general form of $\partial \rho(\textbf{r})/\partial x_i$ are also known. Moreover, this term is responsible for exciting non-zero Fourier modes in $K_i(\textbf{r})$, which allow us to conclude that only the Fourier harmonics present in $\partial \rho(\textbf{r})/\partial x_i$ are excited in the Fourier expansion of $K_{x,y}(\textbf{r})$. In this way, we can propose, without generality loss, that
\begin{equation}
K_i(\textbf{r})=\sum_{m,n} a^{(i)}_{m,n}\sin(k_0\textbf{e}_{m,n}\cdot\textbf{r}),
\end{equation}
where $\textbf{r}=(x\cos(\alpha),y)$ and $\textbf{e}_{m,n}$ is defined as in Eq.~\eqref{bubbles}. Inserting this ansatz for $K_i(\textbf{r})$ and the ground state probability density $\rho(\textbf{r})$ in Eq.~\eqref{action} allows us to obtain upon integration the form of $S[\{a_{m,n}^{(i)}\}]$. The minimization of such many-variables functions gives us finally the set of Fourier amplitudes $\{a_{m,n}^{(i)}\}$ and provide us a direct route to calculate  the components of the superfluid fraction tensor. The main advantage of the method proposed resides in the fact that in most cases a quite limited number of Fourier amplitudes are needed to achieve convergence in the expansions of $K_{x,y}(\textbf{r})$, speeding up significantly the numerical evaluation of the superfluid properties of the system from the knowledge of the ground state wave function.

% We should notice that since the form of $\rho(\textbf{r})$ is known, the symmetry properties and general form of $\partial \rho(\textbf{r})/\partial x_i$ are also known. Moreover, it is exactly this term the one exciting non-zero Fourier modes in $K_i(\textbf{r})$, which allow us to conclude that only the Fourier harmonics present in $\partial \rho(\textbf{r})/\partial x_i$ are present in the Fourier expansion of $K_{x,y}(\textbf{r})$.

\section{Single Mode Results}\label{single_mode_results} 
Upon minimization of the energy per particle functional for the different kinds of solutions considered the ground-state phase diagram of the system is constructed in the density $\rho$ and $a_s/a_{dd}$ plane, for a fixed polarization orientation $\alpha$. Here is worth to remember that the case $\alpha=0$ corresponds to the scenario where polarization is normal to the plane of the system. To understand the main effects of tilting the polarization vector, we first focus on the critical regime of the modulated phases. In this regime, the modulation amplitude is expected to be small and consequently among all the Fourier components in the expansion of $\phi(\textbf{r})$ those corresponding to the first generation of wave vectors are the dominant one. Hence, considering only the leading terms in the Fourier expansion of the different ground-state solutions presented, we obtain the sequence of phase diagrams shown in Fig.~\ref{PhaseDiags} for a frequency trap $\omega=0.08$. As can be observed in Fig.~\ref{PhaseDiags}(a), the phase diagram corresponding to the case in which the polarization is perpendicular to the plane of the sample is reproduced in agreement with previous works \cite{Zhang2021, Ripley2023}. As expected, the phase diagram displays homogeneous, hexagonal, honeycomb, and stripes phases. The honeycomb phase corresponds to a hexagonal solid solution with negative Fourier amplitudes. Moreover, the three lobes corresponding to the modulated phases converge into a critical point, firstly reported by Zhang et al. \cite{Zhang2021} and corrected posteriorly by Ripley et al. \cite{Ripley2023} upon inclusion of the stripes phase. 

Interestingly, as shown in Fig.~\ref{PhaseDiags}(b-f), when the polarization vector is tilted with respect to the plane of the system ($\alpha>0$), the critical point breaks into three critical lines (dashed) ending at critical points (C.P.). One corresponds to a transition from stripes to a homogeneous state, the other corresponds to a transition from a compressed hexagonal phase ($\theta<\pi/6$) to stripes, and a third one is related to a transition from a stretched honeycomb ($\theta>\pi/6$) solution to the stripes phase. The sequence of phase diagrams clearly shows the crossover process in which the critical lines continuously decrease their extension until they converge to a single critical point when $\alpha\rightarrow0$. On the other hand, as the tilting angle $\alpha$ is increased, the extension of the critical lines increases, and the extension of the stripes phase expands significantly, changing completely the behavior of the phase diagram in the region of the critical point for the $\alpha=0$ case. Moreover, for $\alpha>0$, two different triple points (T.P.) are developed in the system, one at low densities where the hexagonal, stripes, and homogeneous phases meet and another at high densities where the honeycomb, stripes and homogeneous phases converge. 

\section{Many modes results}\label{section:many_mode_results} 
\begin{figure}[!t]
    \centering
    \includegraphics[width=0.45\textwidth]{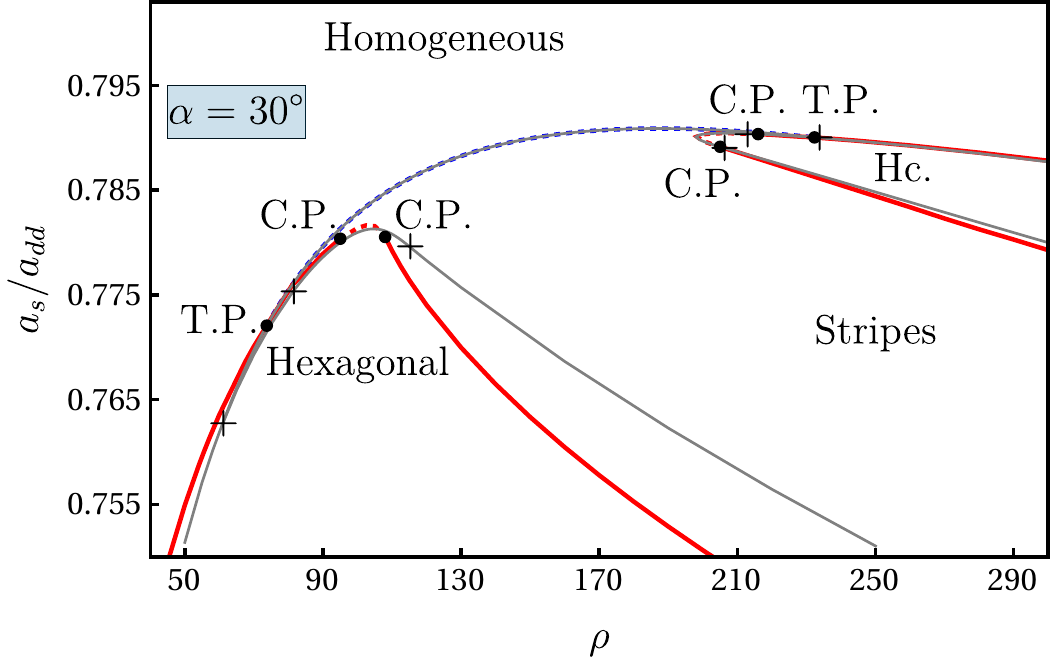}
    \caption{Comparison between the many modes phase diagram and the single mode phase diagrams for a tilting angle $\alpha=30^{\circ}$ and a trap frequency $\omega=0.08$. Red and blue lines correspond to the many modes calculation, while gray lines correspond to the single mode case already presented in Fig.~\ref{PhaseDiags}. The notation employed to name phases and characteristic points is the same followed in Fig.~\ref{PhaseDiags}. Dots (crosses) marks the limits of the critical lines and triple points for the many modes (single mode) case.}
    \label{PD_many_modes}
\end{figure}

Now, we explore the effects of higher-order Fourier modes on the ground state wave function. To this end, we allowed a large enough Fourier basis to achieve a fully converged solution. We choose arbitrarily the case $\alpha=30^\circ$ to compare the many modes and the single mode phase diagrams. As can be observed in Fig.~\ref{PD_many_modes}, the presence of the secondary modes does not alter the stripes-homogeneous critical line (blue dashed). This occurs because higher-order harmonics in the stripes solution become sub-leading at this phase boundary, turning the single mode approximation exact. Meanwhile, the critical lines corresponding to the transitions from compressed hexagonal to stripes phase and from stretched honeycomb to stripes phase are indeed weakly affected by the higher-order harmonics. This occurs because these phase boundaries are located inside the bulk of the stripes phase, which means that, at these phase boundaries, the stripes solution will always contain higher-order harmonics, thus affecting the position of the critical line when higher-order harmonics are included.

\begin{figure}[!t]
    \centering
    \includegraphics[width=0.49\textwidth]{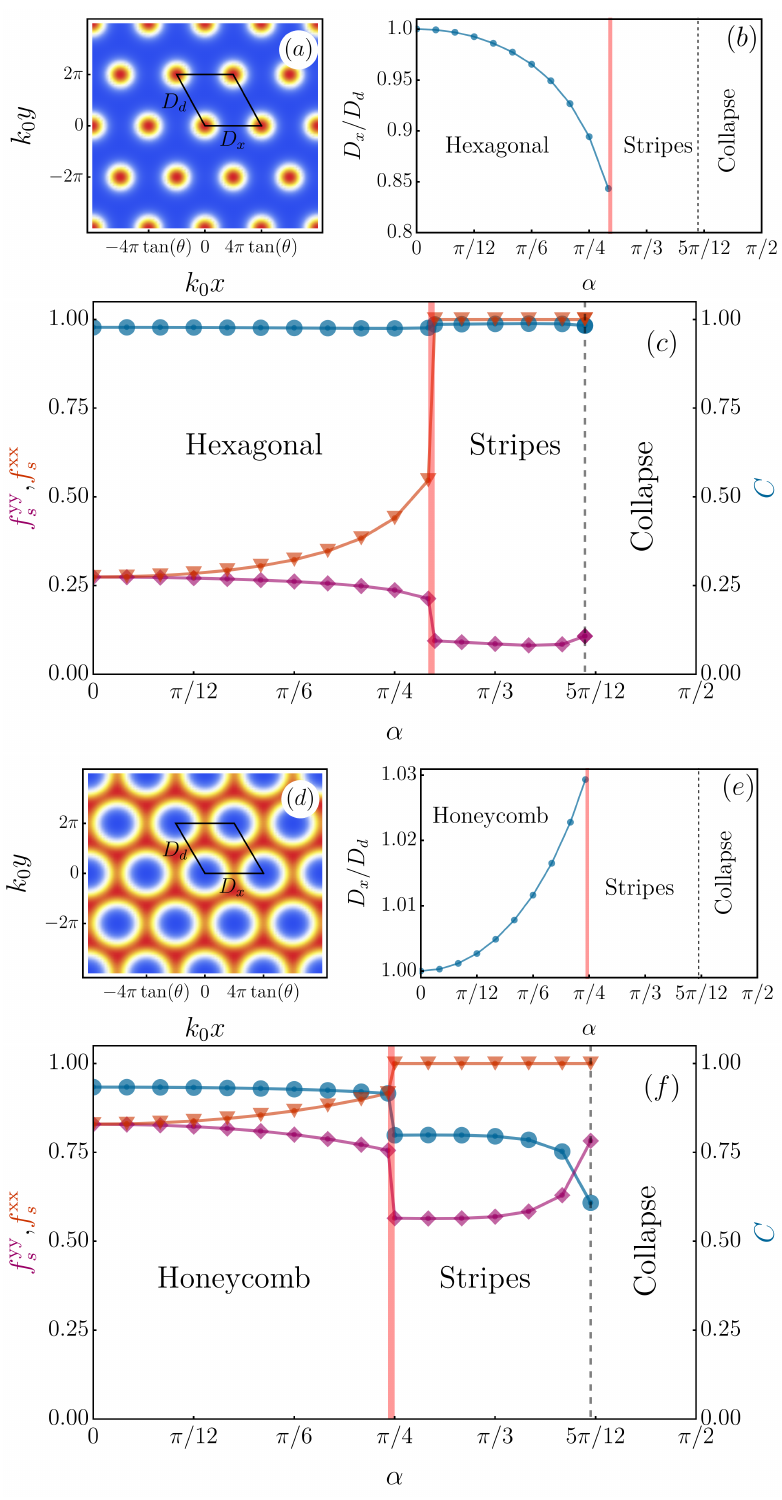}
    \caption{Variation of structural properties with the tilting angle $\alpha$ for the points $(120,0.755)$ (a-c) and $(350,0.779)$ (d-f) in the $(\rho,a_s/a_{dd})$ plane. Figs. (a) and (d) present the density profiles for $\alpha=30^\circ$. Figs. (b) and (e) shows the evolution of the ratio $D_x/D_d$ for the two examples considered. Finally, Figs. (c) and (f) presents the evolution increasing $\alpha$ of the superfluid fraction tensor components $f_s^{xx}$ (orange triangles) and $f_s^{yy}$ (purple diamonds) and the contrast ($C$) (blue circles) for the two example points considered. The pink stripe localizing the transition between modulated phases in (c) and (f)  has a width equal to $1^\circ$. %{\color{red}In my opinion, the figures showing the superfluid fraction and contrast could have different markers for each quantity, not only different colors. Each color and marker relating to each quantity has to be in the caption of the figure to facilitate the interpretation of it.}\textcolor{blue}{(Figures (c) and (f) need a distinct marker for each quantity for better visibility and interpretation. Maybe the pink stripe needs to be a different color for better readability of the lines.)}
    }
    \label{Struct_Prop}
\end{figure}

Despite the previous discussion, a comparison between Fig.~\ref{PhaseDiags} and Fig.~\ref{PD_many_modes} shows that the critical lines, their ending points, and the triple points were not strongly affected by the single mode approximation in this case. The most significant difference takes place in the position of the first-order transition between the hexagonal and the stripes phases deep into the modulated regime. The stability enhancement of the stripes is somewhat expected since this is the kind of modulated solution most affected by the single-mode approximation, while the hexagonal solution contains two independent variational Fourier amplitudes, at this level of approximation, the stripes solution contains only one.

Now we turn our attention to the impact on the structural properties of tilting the polarization of the system. As already mentioned in Sec.~\ref{single_mode_results}, the phases with hexagonal symmetry for $\alpha=0$ are deformed when $\alpha>0$. In general, the presence of a tilted polarization with respect to the plane of the system creates an additional effective attractive dipolar interaction between the clusters of particles in the direction of the in-plane polarization. This leads to a compression of the hexagonal solid phase and a stretching of the lattice of voids in the honeycomb phase along the $x$-direction. This effect can be measured by the ratio between the distance between the next nearest clusters (voids) along the $x$-direction ($D_x$) and along the diagonal direction ($D
_d$). It is not hard to show that given the anzats for $\phi_{\mathrm{hx}}(\textbf{r})$, we have that $D_x/D_d=2\sin(\theta)$. Here $D_x$ and $D_d$ are measured between the center of clusters (voids) and on a plane perpendicular to the polarization vector, which sets the 3D filaments orientation.

In Fig.~\ref{Struct_Prop}(a-b) and  Fig.~\ref{Struct_Prop}(d-e) we show two examples of the probability density patterns $\phi^2(\textbf{r})$ for two different points in the $(\rho,a_s/a_{dd})$ plane for $\alpha=30^{\circ}$, the definition of $D_x$ and $D_d$ and the behavior of the corresponding ratio $D_x/D_{d}$ varying $\alpha$. We can observe that, for the example values considered, the deformation of the hexagonal phase is much greater than the one experienced by the honeycomb phase as the polarization tilting angle $\alpha$ is increased. This seems to be reflecting a higher resistance to deformations of the honeycomb in comparison with the hexagonal pattern. Such deformations of the solutions not only affect the geometric features of the density pattern but, more interestingly, they impact the superfluid properties of the system.  

As discussed in the literature, long wavelength properties in a system with hexagonal symmetry are typically isotropic, this occurs for instance with the elastic response and also occurs with the superfluid fraction tensor~\cite{Landau_1986,Blakie_2024}. In this sense, a natural check of the technique employed to compute the superfluid fraction tensor is to recover the same values for $f_s^{xx}$ and $f_s^{yy}$ for the hexagonally symmetric configurations obtained when $\alpha=0$. As can be observed in Fig.~\ref{Struct_Prop}(c) and Fig.~\ref{Struct_Prop}(f), the superfluid tensor components along $x$ and $y$ directions are indeed equal for perpendicular polarization ($\alpha=0$). Moreover, the off-diagonal components of the superfluid fraction tensor remain zero for all $\alpha$'s. A natural result, since both hexagonally symmetric patterns and stripes patterns have zero off-diagonal components when the configuration is symmetric against reflections with respect to the $x$ and $y$ axes. To complement our discussion of the superfluid properties we simultaneously present in Fig.~\ref{Struct_Prop}(c) and Fig.~\ref{Struct_Prop}(f) the contrast of the corresponding modulated phases, defined as \begin{equation}
C=\frac{\mathrm{Max}(\vert\psi\vert^2)-\mathrm{Min}(\vert\psi\vert^2)}{\mathrm{Max}(\vert\psi\vert^2)+\mathrm{Min}(\vert\psi\vert^2)}.
\end{equation}
The results show that as the polarization develops a component along the $x$-direction ($\alpha>0$) the superfluid fraction along this direction ($f_s^{xx}$) grows steadily while the $y$ component decreases as long as we remain within the hexagonal and honeycomb phases. In the case of the hexagonal phase this is a response to the tendency of the system of approximate clusters along the $x$-direction and make them further apart in the other directions when $\alpha$ is increased. Interestingly, despite the significant variation of the superfluid properties, the contrast of the modulated phases does not display a significant variation along this process.

In the case of the honeycomb phase, a smaller anisotropy between $f_s^{xx}$ and $f_s^{yy}$ is observed when compared with the hexagonal case. In this scenario, the leading mechanism for the variation of the superfluidity as $\alpha$ is increased does not seem to be related to the deformation of the honeycomb lattice since it remains at small values over the whole phase (see Fig.~\ref{Struct_Prop}(e)). The variation of the superfluidity with $\alpha$ in this case is produced by a process of particle redistribution in which the regions linking maximum density points along the $x$-direction in a ``zig-zag" trajectory (see Fig.~\ref{Struct_Prop}(d)) gets more homogeneous and wider while regions linking maxima along the $y$-direction becomes more localized. 

% In the case of the honeycomb phase, a smaller anisotropy between $f_s^{xx}$ and $f_s^{yy}$ is observed when compared with the hexagonal case. In this scenario, the leading mechanism for the variation of the superfluidity increasing $\alpha$ does not seem to be related to the deformation of the honeycomb lattice since it remains at small values over the whole phase (see Fig.\ref{Struct_Prop} (e)).

Within the stripes region we have $f_s^{xx}(\alpha)=1$, while $f_s^{yy}(\alpha)$ develops in general a non monotonic behavior (see Fig.~\ref{Struct_Prop}(c)). The local monotony of  $f_s^{yy}(\alpha)$ is defined by the interplay of two competing effects as $\alpha$ is increased. On the one hand, increasing $\alpha$ favors dipolar attraction between particles in a given stripe, promoting localization which hinders superfluidity between stripes. On the other hand, the total dipolar field localizing dipoles in a given stripe depends on the width ($2\sigma$) of the column along the $z$-direction, which decreases when the attraction between single dipoles increases, i.e., when we increase $\alpha$. The dominant effect between these two will define the monotony of the superfluid fraction curve $f_s^{yy}(\alpha)$. Our results seem to indicate that, for large enough polarization tilting, the decrease of $\sigma$ is always strong enough to produce a softening of the modulated pattern, increasing transverse superfluidity. Finally, we also identify in Fig.~\ref{Struct_Prop} the region at which the total two-body effective interaction at zero momentum becomes negatives, i.e. $\hat{V}_{\mathrm{eff}}(0)+a_s/(5a_{dd})<0$. This region is identified as the collapse region, following established literature for strictly 2D dipolar systems~\cite{Rag_2015,Mishra_2016,Shen_2021}. 

% In the other hand, within the stripes region $f_s^{xx}(\alpha)=1$ while $f_s^{yy}(\alpha)$ develops in general a non monotonic behavior (see Fig.~\ref{Struct_Prop}(c)). The local monotony of  $f_s^{yy}(\alpha)$ is defined by the interplay of two competing effects as $\alpha$ is increased. On one hand, the increasing of $\alpha$ increases the dipolar attraction between particles in a given stripe, favoring localization and hence increasing the modulation amplitude, which hinders superfluidity between stripes. On the other hand, the total dipolar field localizing dipoles in a given stripe depends on the width ($2\sigma$) of the column along the $z$-direction, which decreases when the attraction between single dipoles increases, i.e. when we increase $\alpha$. The dominant effect between these two will define the monotony of the superfluid fraction curve $f_s^{yy}(\alpha)$. Our results seem to indicate that, for large enough polarization tilting, the decrease of $\sigma$ is always strong enough to produce a softening of the modulated pattern, increasing transverse superfluidity. Finally, we also identify in Fig.\ref{Struct_Prop} the region at which the total two-body effective interaction at zero momentum becomes negatives, $\hat{V}_{\mathrm{eff}}(0)+a_s/(5a_{dd})<0$. This region is identified as the collapse region, following established literature for strictly 2D dipolar systems~\cite{Rag_2015,Mishra_2016,Shen_2021}. 

\section{Final Discussion}
In the last few years, a surge of interest in the possible impacts of topological ingredients on Bose-Einstein condensation has emerged as a hot topic in the search for exotic phases in the field of ultracold quantum gases. Different authors have considered bubble, torus, cylinder among other possible geometries to study how different configurations produce non-trivial condensate properties and modulated phases~\cite{Tononi_2019, Tononi_2020, Tononi_2023, Young_2023, Ciardi_2024}. In this sense, we consider in the present work a relatively simple setup that has remained unexplored so far \cite{Tanzi2019, Lu2011, Aikawa2012}, a planar polarized dipolar Bose gas with tilted polarization with respect to the plane of the system. Our results show that tilting the polarization orientation has a major impact on the quantum critical behavior of the system when compared with the already known case with perpendicular polarization. In this case, as the polarization angle departs from the normal direction the isolated quantum critical point present at $\alpha=0$ breaks into three critical lines separating two phases at a time. Besides changing the critical behavior, a tilted polarization also produces structural changes in the modulated phases, mainly in the hexagonal and honeycomb phases, which develop axially anisotropic properties. In this respect, we obtain that this axial anisotropy induced by the in-plane polarization favors superfluidity along this direction and hinders it in the orthogonal direction. This result offers an interesting avenue to manipulate the superfluid properties of hexagonal and honeycomb dipolar supersolids, which otherwise ($\alpha=0$) present an isotropic behavior.  

On another subject, it is important to remark that despite the existence of significant literature exploring the physics of strictly 2D dipolar systems with tilted polarization, the physical behavior of the analog quasi-2D systems is radically different and significantly under researched. Only recently the ground-state phase diagram of quasi-2D dipolar systems with perpendicular polarization was established \cite{Ripley2023}, and the effects of the tilted polarization concerning the plane of the system remain unexplored until now.

We conclude by discussing the important aspect of the experimental realizations. The regime of parameters used for analytical calculations is compatible with current experimental capabilities~\cite{Chomaz2018, Kadau2016, Schmitt2016, Tanzi2019, Bottcher2019}. A potential experiment using $^{162}$Dy would allow a wide range of $s$-wave scattering lengths $a_s$. Considering the dipolar length $a_{dd}\approx 7\,\mathrm{nm}$, we will have a range of $a_s/a_{dd}$ consistent with the values considered in this work. For $^{162}$Dy, the characteristic units of length and time will be $\ell=0.26\,\mathrm{\mu m}$ and 
$t_0=0.18\,\mathrm{ms}$. Hence a trapping dimensionless frequency $\omega_z\, t_0=0.08$ is equivalent to  
$\omega_z\approx450\,\mathrm{Hz}$. As an example, let us consider the configuration presented in Fig.~\ref{Struct_Prop}(a) corresponding to $a_s/a_{dd}=0.755$, $\rho=120$ and $\alpha=30^\circ$, in this case, we will have $\sigma \approx6.78\,\mathrm{\mu m}$ and a peak $3$D density along the middle plane of the system $3\rho/4\sigma\approx1.9\times10^{14}\,\mathrm{cm^{-3}}$. For these conditions, the ground-state characteristic wave vector of the compressed hexagonal solid is $k_0\approx0.4$, which results in a lattice spacing $D_x=4\pi\tan(\theta)/k_0\approx 4.59\,\mathrm{\mu m}$ and $D_d=2\pi\sec(\theta)/k_0\approx 4.75\,\mathrm{\mu m}$, i.e., a sufficiently small value that allows 
studying long-distance physical properties experimentally.

\begin{acknowledgments} A.M.C. acknowledge UNIFI for financial support and hospitality. A.M.C., D.L., and M.G. acknowledge the Fundação de Amparo à Pesquisa de Santa Catarina, Brazil (Fapesc), and Coordenação de Aperfeiçoamento de Pessoal de Nível Superior - Brasil (CAPES) - Finance Code 001. Fapesc, CAPES, and CNPq supported this work. F. C. and V. Z. acknowledge financial support from PNRR MUR Project No. PE0000023-NQSTI. 
\end{acknowledgments}

\bibliography{dipolar.bib}

\end{document}